\documentclass{elsart}

\usepackage{graphicx}

\newcommand{\be}{\begin{equation}}
\newcommand{\ee}{\end{equation}}
\newcommand{\beq}{\begin{eqnarray}}
\newcommand{\eeq}{\end{eqnarray}}
\newcommand{\mbf}{\mathbf}
\newcommand{\ds}{\displaystyle}

\begin{document}
\begin{frontmatter}
\title{Granger Causality: Basic Theory and Application to
Neuroscience}

\author{Mingzhou Ding, Yonghong Chen}
\address{The J. Crayton Pruitt Family Department of Biomedical Engineering, University of Florida, USA}

\author{Steven~L. Bressler}
\address{Center for Complex Systems and Brain Sciences, Florida Atlantic University, USA}

\newpage
\begin{abstract}

Multi-electrode neurophysiological recordings produce massive
quantities of data. Multivariate time series analysis provides the
basic framework for analyzing the patterns of neural interactions in
these data. It has long been recognized that neural interactions are
directional. Being able to assess the directionality of neuronal
interactions is thus a highly desired capability for understanding
the cooperative nature of neural computation. Research over the last
few years has shown that Granger causality is a key technique to
furnish this capability. The main goal of this article is to provide
an expository introduction to the concept of Granger causality.
Mathematical frameworks for both bivariate Granger causality and
conditional Granger causality are developed in detail with
particular emphasis on their spectral representations. The technique
is demonstrated in numerical examples where the exact answers of
causal influences are known.  It is then applied to analyze
multichannel local field potentials recorded from monkeys performing
a visuomotor task. Our results are shown to be physiologically
interpretable and yield new insights into the dynamical organization
of large-scale oscillatory cortical networks.
\end{abstract}
\end{frontmatter}
\newpage

\section{Introduction}
In neuroscience, as in many other fields of science and engineering,
signals of interest are often collected in the form of multiple
simultaneous time series. To evaluate the statistical
interdependence among these signals, one calculates cross
correlation functions in the time domain and ordinary coherence
functions in the spectral domain. However, in many situations of
interest, symmetric\footnote{Here by symmetric we mean that, when A
is coherent with B, B is equally coherent with A.} measures like
ordinary coherence are not completely satisfactory, and further
dissection of the interaction patterns among the recorded signals is
required to parcel out effective functional connectivity in complex
networks. Recent work has begun to consider the causal influence one
neural time series can exert on another. The basic idea can be
traced back to Wiener \cite{wiener} who conceived the notion that,
if the prediction of one time series could be improved by
incorporating the knowledge of a second one, then the second series
is said to have a causal influence on the first. Wiener's idea lacks
the machinery for practical implementation. Granger later formalized
the prediction idea in the context of linear regression models
\cite{granger}. Specifically, if the variance of the autoregressive
prediction error of the first time series at the present time is
reduced by inclusion of past measurements from the second time
series, then the second time series is said to have a causal
influence on the first one. The roles of the two time series can be
reversed to address the question of causal influence in the opposite
direction. From this definition it is clear that the flow of time
plays a vital role in allowing inferences to be made about
directional causal influences from time series data. The interaction
discovered in this way may be reciprocal or it may be
unidirectional.

Two additional developments of Granger's causality idea are
important. First, for three or more simultaneous time series, the
causal relation between any two of the series may be direct, may
be mediated by a third one, or may be a combination of both. This
situation can be addressed by the technique of conditional Granger
causality. Second, natural time series, including ones from
economics and neurobiology, contain oscillatory aspects in
specific frequency bands. It is thus desirable to have a spectral
representation of causal influence. Major progress in this
direction has been made by Geweke \cite{geweke,geweke2} who found
a novel time series decomposition technique that expresses the
time domain Granger causality in terms of its frequency content.
In this article we review the essential mathematical elements of
Granger causality with special emphasis on its spectral
decomposition. We then discuss practical issues concerning how to
estimate such measures from time series data. Simulations are used
to illustrate the theoretical concepts. Finally, we apply the
technique to analyze the dynamics of a large-scale sensorimotor
network in the cerebral cortex during cognitive performance. Our
result demonstrates that, for a well designed experiment, a
carefully executed causality analysis can reveal insights that are
not possible with other techniques.

\section{Bivariate Time Series and Pairwise Granger Causality}

Our exposition in this and the next section follows closely that of
Geweke \cite{geweke,geweke2}. To avoid excessive mathematical
complexity we develop the analysis framework for two time series.
The framework can be generalized to two sets of time series
\cite{geweke}.

\subsection{Time Domain Formulation}

Consider two stochastic processes $X_t$ and $Y_t$. Assume that they
are jointly stationary. Individually, under fairly general
conditions, each process admits an autoregressive
representation\index{autoregressive model} \be \label{xsingle}
X_t=\sum^{\infty}_{j=1} a_{1j} X_{t-j}+\epsilon_{1t},\ \
\rm{var}(\epsilon_{1t})=\Sigma_1. \ee \be \label{ysingle}
Y_t=\sum^{\infty}_{j=1} d_{1j} Y_{t-j}+\eta_{1t},\ \
\rm{var}(\eta_{1t})=\Gamma_1. \ee Jointly, they are represented as
\be \label{xdouble} X_t=\sum^{\infty}_{j=1} a_{2j}
X_{t-j}+\sum^{\infty}_{j=1} b_{2j} Y_{t-j}+\epsilon_{2t}, \ee \be
\label{ydouble} Y_t=\sum^{\infty}_{j=1} c_{2j}
X_{t-j}+\sum^{\infty}_{j=1} d_{2j} Y_{t-j}+\eta_{2t},\ee where the
noise terms are uncorrelated over time and their contemporaneous
covariance matrix is
\beq {\mbf \Sigma} = \left(%
\begin{array}{cc}
  \Sigma_2 \ \ & \Upsilon_2 \\
  \Upsilon_2 \ \ & \Gamma_2 \\
\end{array}%
\right).\eeq The entries are defined as
$\Sigma_2=\rm{var}(\epsilon_{2t}), \Gamma_2=\rm{var}(\eta_{2t}),
\Upsilon_2=cov(\epsilon_{2t},\eta_{2t})$.  If $X_t$ and $Y_t$ are
independent, then $\{b_{2j}\}$ and $\{c_{2j}\}$ are uniformly zero,
$\Upsilon_2=0$, $\Sigma_1=\Sigma_2$ and $\Gamma_1=\Gamma_2$. This
observation motivates the definition of total
interdependence\index{interdependence} between $X_t$ and $Y_t$ as
\be \label{total} F_{X,Y}=\ln \frac{\Sigma_1 \Gamma_1}{|{\mbf
\Sigma}|} \ee where $|\cdot|$ denotes the determinant of the
enclosed matrix. According to this definition, $F_{X,Y}=0$ when the
two time series are independent, and $F_{X,Y}>0$ when they are not.

Consider Eqs. (\ref{xsingle}) and (\ref{xdouble}). The value of
$\Sigma_1$ measures the accuracy of the autoregressive prediction of
$X_t$ based on its previous values, whereas the value of $\Sigma_2$
represents the accuracy of predicting the present value of $X_t$
based on the previous values of both $X_t$ and $Y_t$. According to
Wiener \cite{wiener} and Granger \cite{granger}, if $\Sigma_2$ is
less than $\Sigma_1$ in some suitable statistical sense, then $Y_t$
is said to have a causal influence on $X_t$. We quantify this causal
influence\index{causal influence} by \be F_{Y\rightarrow X}=\ln
\frac{\Sigma_1}{\Sigma_2}. \ee It is clear that $F_{Y\rightarrow
X}=0$ when there is no causal influence from $Y$ to $X$ and
$F_{Y\rightarrow X}>0$ when there is. Similarly, one can define
causal influence\index{causal influence} from $X$ to $Y$ as \be
F_{X\rightarrow Y}=\ln \frac{\Gamma_1}{\Gamma_2}. \ee

It is possible that the interdependence between $X_t$ and $Y_t$
cannot be fully explained by their interactions. The remaining
interdependence is captured by $\Upsilon_2$, the covariance between
$\epsilon_{2t}$ and $\eta_{2t}$. This interdependence is referred to
as instantaneous causality\index{instantaneous causality} and is
characterized by \be F_{X \cdot Y}=\ln
\frac{\Sigma_2\Gamma_2}{\left|\mbf \Sigma\right|}. \ee When
$\Upsilon_2$ is zero, $F_{X \cdot Y}$ is also zero. When
$\Upsilon_2$ is not zero, $F_{X \cdot Y}>0$.

The above definitions imply that \be \label{relationship}
F_{X,Y}=F_{X \rightarrow Y}+F_{Y \rightarrow X}+F_{X \cdot Y}. \ee
Thus we decompose the total interdependence between two time
series $X_t$ and $Y_t$ into three components: two directional
causal influences due to their interaction patterns, and the
instantaneous causality due to factors possibly exogenous to the
$(X,Y)$ system (e.g. a common driving input).

\subsection{Frequency Domain Formulation}

To begin we define the lag operator $L$ to be $LX_t=X_{t-1}$.
Rewrite Eqs. (\ref{xdouble}) and (\ref{ydouble}) in terms of the lag
operator
\beq \label{xydouble} \left(%
\begin{array}{cc}
  a_2(L) \ \ & b_2(L) \\
  c_2(L) \ \ & d_2(L) \\
\end{array}%
\right)\left(%
\begin{array}{c}
  X_t \\
  Y_t \\
\end{array}%
\right) =\left(%
\begin{array}{c}
  \epsilon_{2t} \\
  \eta_{2t} \\
\end{array}%
\right), \eeq where $a_2(0)=1, b_2(0)=0, c_2(0)=0, d_2(0)=1$.
Fourier transforming both sides of Eq. (\ref{xydouble}) leads to
\beq \label{xyspectrum} \left(%
\begin{array}{cc}
  a_2(\omega) \ \ & b_2(\omega) \\
  c_2(\omega) \ \ & d_2(\omega) \\
\end{array}%
\right)\left(%
\begin{array}{c}
  X(\omega) \\
  Y(\omega) \\
\end{array}%
\right) =\left(%
\begin{array}{c}
  E_x(\omega) \\
  E_y(\omega) \\
\end{array}%
\right), \eeq where the components of the coefficient matrix $\mbf
A(\omega)$ are \beq a_2(\omega)=1-\sum_{j=1}^{\infty}
a_{2j}e^{-i\omega j},\ \
b_2(\omega)=-\sum_{j=1}^{\infty} b_{2j}e^{-i\omega j},\nonumber \\
c_2(\omega)=-\sum_{j=1}^{\infty} c_{2j}e^{-i\omega j},\ \
d_2(\omega)=1-\sum_{j=1}^{\infty} d_{2j}e^{-i\omega j}. \nonumber
\eeq Recasting Eq. (\ref{xyspectrum}) into the transfer function
format we obtain \beq \label{xyspm}
\left(%
\begin{array}{c}
  X(\omega) \\
  Y(\omega) \\
\end{array}%
\right) =  \left(%
\begin{array}{cc}
  H_{xx}(\omega) \ \ & H_{xy}(\omega) \\
  H_{yx}(\omega) \ \ & H_{yy}(\omega) \\
\end{array}%
\right)\left(%
\begin{array}{c}
  E_x(\omega) \\
  E_y(\omega) \\
\end{array}%
\right), \eeq where the transfer function is $\mbf H(\omega)=\mbf
A^{-1}(\omega)$ whose components are\beq \label{trans1}
H_{xx}(\omega)=\frac{1}{{\rm det}\mbf A}d_2(\omega),\ \
H_{xy}(\omega)=-\frac{1}{{\rm det}\mbf A}b_2(\omega), \nonumber
\\ H_{yx}(\omega)=-\frac{1}{{\rm det}\mbf A}c_2(\omega), \ \
H_{yy}(\omega)=\frac{1}{{\rm det}\mbf A}a_2(\omega). \eeq After
proper ensemble averaging we have the spectral matrix \be
\label{spectralmatrix} {\mbf S}(\omega)={\mbf H}(\omega){\mbf
\Sigma}{\mbf H}^*(\omega) \ee where * denotes complex conjugate and
matrix transpose.

The spectral matrix contains cross spectra and auto spectra. If
$X_t$ and $Y_t$ are independent, then the cross spectra are zero and
$|{\mbf S}(\omega)|$ equals the product of two auto spectra. This
observation motivates the spectral domain representation of total
interdependence\index{interdependence} between $X_t$ and $Y_t$ as
\beq \label{totaldecompose} f_{X,Y}(\omega) = \ln
\frac{S_{xx}(\omega)S_{yy}(\omega)} {|\mbf S(\omega)|}, \eeq
 where $|\mbf
 S(\omega)|=S_{xx}(\omega)S_{yy}(\omega)-S_{xy}(\omega)S_{yx}(\omega)$ and
 $S_{yx}(\omega)=S_{xy}^*(\omega)$.
It is easy to see that this decomposition of interdependence is
related to coherence by the following relation: \be
\label{cohrelation} f_{X,Y}(\omega) = -\ln (1-C(\omega)), \ee where
coherence\index{coherence} is defined as $$C(\omega) = \displaystyle
\frac{|S_{xy}(\omega)|^2}{S_{xx}(\omega)S_{yy}(\omega)}.$$ The
coherence defined in this way is sometimes referred to as the
squared coherence.

To obtain the frequency decomposition of the time domain causality
defined in the previous section, we look at the auto spectrum of
$X_t$: \be \label{xspectrum} S_{xx}(\omega)=
H_{xx}(\omega)\Sigma_2 H_{xx}^*(\omega) + 2\Upsilon_2 {\rm Re}
(H_{xx}(\omega) H_{xy}^*(\omega)) + H_{xy}(\omega)\Gamma_2
H_{xy}^*(\omega).\ee It is instructive to consider the case where
$\Upsilon_2=0$. In this case there is no instantaneous causality
and the interdependence between $X_t$ and $Y_t$ is entirely due to
their interactions through the regression terms on the right hand
sides of Eqs. (\ref{xdouble}) and (\ref{ydouble}). The spectrum
has two terms. The first term, viewed as the intrinsic part,
involves only the variance of $\epsilon_{2t}$, which is the noise
term that drives the $X_t$ time series. The second term, viewed as
the causal part, involves only the variance of $\eta_{2t}$, which
is the noise term that drives $Y_t$. This power decomposition into
an ``intrinsic'' term and a ``causal" term will become important
for defining a measure for spectral domain causality.

When $\Upsilon_2$ is not zero it becomes harder to attribute the
power of the $X_t$ series to different sources. Here we consider a
transformation introduced by Geweke \cite{geweke} that removes the
cross term and makes the identification of an intrinsic power term
and a causal power term possible. The procedure is called
normalization and it consists of left-multiplying
\be \label{trans2} \mbf P=\left(%
\begin{array}{cc}
  1 \ \ & 0 \\
  -\ds{\frac{\Upsilon_2}{\Sigma_2}} \ \ & 1 \\
\end{array}%
\right)\ee on both sides of Eq. (\ref{xyspectrum}). The result is \beq \label{xtransf} \left(%
\begin{array}{cc}
  a_2(\omega) & b_2(\omega) \\
  c_3(\omega) & d_3(\omega) \\
\end{array}%
\right) \left(%
\begin{array}{c}
  X(\omega) \\
  Y(\omega) \\
\end{array}%
\right)= \left(%
\begin{array}{c}
  E_x(\omega) \\
  \tilde{E} _y(\omega)\\
\end{array}%
\right), \eeq where
$c_3(\omega)=c_2(\omega)-\ds{\frac{\Upsilon_2}{\Sigma_2}}
a_2(\omega),
d_3(\omega)=d_2(\omega)-\ds{\frac{\Upsilon_2}{\Sigma_2}}
b_2(\omega),
\tilde{E}_y(\omega)=E_y(\omega)-\ds{\frac{\Upsilon_2}{\Sigma_2}}
E_x(\omega)$. The new transfer function $\tilde{\mbf H}(\omega)$ for
(\ref{xtransf}) is the inverse of the new coefficient matrix
$\tilde{\mbf A}(\omega)$: \be \label{newH}
 \tilde{\mbf H}(\omega)=\left(%
\begin{array}{cc}
  \tilde{H}_{xx}(\omega) \ \ & \tilde{H}_{xy}(\omega) \\
  \tilde{H}_{yx}(\omega) \ \ & \tilde{H}_{yy}(\omega) \\
\end{array}%
\right)=\frac{1}{\det \tilde{\mbf A}} \left(%
\begin{array}{cc}
  d_3(\omega) \ \ & -b_2(\omega) \\
  -c_3(\omega) \ \ & a_2(\omega) \\
\end{array}%
\right).\ee Since $\det \tilde{\mbf A}=\det \mbf A$ we have \beq
\tilde{H}_{xx}(\omega)=H_{xx}(\omega)+\ds{\frac{\Upsilon_2}{\Sigma_2}}H_{xy}(\omega),\
\
\tilde{H}_{xy}(\omega)=H_{xy}(\omega),\nonumber \\
\tilde{H}_{yx}(\omega)=H_{yx}(\omega)+\ds{\frac{\Upsilon_2}{\Sigma_2}}H_{xx}(\omega),\
\ \tilde{H}_{yy}(\omega)=H_{yy}(\omega). \eeq From the construction
it is easy to see that $E_x$ and $\tilde{E}_y$ are uncorrelated,
that is, $\rm{cov}(E_x,\tilde{E}_y)=0$. The variance of the noise
term for the normalized $Y_t$ equation is
$\tilde{\Gamma}_2=\Gamma_2-\ds{\frac{\Upsilon^2_2}{\Sigma_2}}$. From
Eq. (\ref{xtransf}), following the same steps that lead to Eq.
(\ref{xspectrum}), the spectrum of $X_t$ is found to be: \be
\label{newxspm} S_{xx}(\omega)=\tilde{H}_{xx}(\omega)\Sigma_2
\tilde{H}_{xx}^*(\omega) + H_{xy}(\omega)\tilde{\Gamma}_2
H_{xy}^*(\omega).\ee Here the first term is interpreted as the
intrinsic power and the second term as the causal power of $X_t$ due
to $Y_t$. This is an important relation because it explicitly
identifies that portion of the total power of $X_t$ at frequency
$\omega$ that is contributed by $Y_t$. Based on this interpretation
we define the causal influence\index{causal influence} from $Y_t$ to
$X_t$ at frequency $\omega$ as \be \label{xmeasure} f_{Y\rightarrow
X}(\omega)=\ln \frac{S_{xx}(\omega)}{\tilde{H}_{xx}(\omega)\Sigma_2
\tilde{H}_{xx}^*(\omega)}.\ee Note that this definition of causal
influence is expressed in terms of the intrinsic power rather than
the causal power. It is expressed in this way so that the causal
influence is zero when the causal power is zero (i.e., the intrinsic
power equals the total power), and increases as the causal power
increases (i.e., the intrinsic power decreases).

By taking the transformation matrix as $\left(%
\begin{array}{cc}
  1 & -\Upsilon_2/\Gamma_2 \\
  0 & 1 \\
\end{array}%
\right)$ and performing the same analysis, we get the causal
influence\index{causal influence} from $X_t$ to $Y_t$: \be
\label{ymeasure} f_{X\rightarrow Y}(\omega)=\ln
\frac{S_{yy}(\omega)}{\hat{H}_{yy}(\omega)\Gamma_2
\hat{H}_{yy}^*(\omega)},\ee where
$\hat{H}_{yy}(\omega)=H_{yy}(\omega)+\ds{\frac{\Upsilon_2}{\Gamma_2}}
H_{yx}(\omega)$.

By defining the spectral decomposition of instantaneous
causality\index{instantaneous causality} as \cite{gm} \be
\label{measureinst} f_{X \cdot Y}(\omega)=\ln
\frac{(\tilde{H}_{xx}(\omega)\Sigma_2
\tilde{H}_{xx}^*(\omega))(\hat{H}_{yy}(\omega)\Gamma_2
\hat{H}_{yy}^*(\omega))}{|\mbf S(\omega)|},\ee we achieve a spectral
domain expression for the total interdependence that is analogous to
Eq. (\ref{relationship}) in the time domain, namely: \be
\label{freqrelationship} f_{X,Y}(\omega)=f_{X \rightarrow
Y}(\omega)+f_{Y \rightarrow X}(\omega)+f_{X \cdot Y}(\omega). \ee We
caution that the spectral instantaneous causality may become
negative for some frequencies in certain situations and may not have
a readily interpretable physical meaning.

It is important to note that, under general conditions, these
spectral measures relate to the time domain measures as:\beq F_{Y,X}
& = & \frac{1}{2\pi}\int_{-\pi}^{\pi} f_{Y,
X}(\omega)d \omega,\nonumber \\
F_{Y\rightarrow X} & = & \frac{1}{2\pi}\int_{-\pi}^{\pi}
f_{Y\rightarrow
X}(\omega)d \omega,\nonumber \\
F_{X\rightarrow Y} & = & \frac{1}{2\pi}\int_{-\pi}^{\pi}
f_{X\rightarrow Y}(\omega)d \omega,\nonumber \\ F_{Y \cdot X} & =
& \frac{1}{2\pi}\int_{-\pi}^{\pi} f_{Y \cdot X}(\omega)d
\omega.\eeq The existence of these equalities gives credence to
the spectral decomposition procedures described above.

\section{Trivariate Time Series and Conditional Granger Causality}

\begin{figure}
\begin{center}
\includegraphics[scale=0.8]{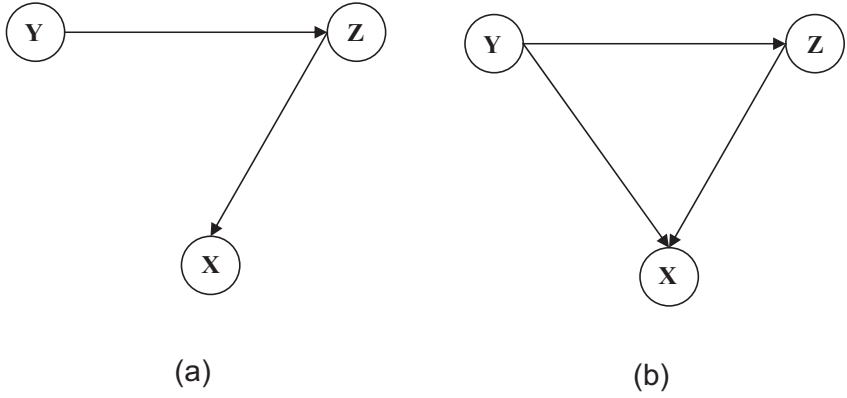}
\end{center}
\caption{Two distinct patterns of connectivity among three time
series. A pairwise causality analysis cannot distinguish these two
patterns.}
     \label{figure1}
     \end{figure}

For three or more time series one can perform a pairwise analysis
and thus reduce the problem to a bivariate problem. This approach
has some inherent limitations. For example, for the two coupling
schemes in Figure \ref{figure1}, a pairwise analysis will give the
same patterns of connectivity like that in Figure \ref{figure1}(b).
Another example involves three processes where one process drives
the other two with differential time delays. A pairwise analysis
would indicate a causal influence from the process that receives an
early input to the process that receives a late input. To
disambiguate these situations requires additional measures. Here we
define conditional Granger causality which has the ability to
resolve whether the interaction between two time series is direct or
is mediated by another recorded time series and whether the causal
influence is simply due to differential time delays in their
respective driving inputs. Our development is for three time series.
The framework can be generalized to three sets of time series
\cite{geweke2}.

\subsection{Time Domain Formulation}

Consider three stochastic processes $X_t$, $Y_t$ and $Z_t$. Suppose
that a pairwise analysis reveals a causal influence from $Y_t$ to
$X_t$. To examine whether this influence has a direct component
(Figure \ref{figure1}(b)) or is mediated entirely by $Z_t$ (Figure
\ref{figure1}(a)) we carry out the following procedure. First, let
the joint autoregressive representation of $X_t$ and $Z_t$ be \be
\label{xcdouble} X_t=\sum^{\infty}_{j=1} a_{3j}
X_{t-j}+\sum^{\infty}_{j=1} b_{3j} Z_{t-j}+\epsilon_{3t}, \ee \be
\label{zcdouble} Z_t=\sum^{\infty}_{j=1} c_{3j}
X_{t-j}+\sum^{\infty}_{j=1} d_{3j} Z_{t-j}+\gamma_{3t},\ee where the
covariance matrix of the noise terms is
\beq {\mbf \Sigma}_3 = \left(%
\begin{array}{cc}
  \Sigma_3 \ \ & \Upsilon_3 \\
  \Upsilon_3 \ \ & \Gamma_3 \\
\end{array}%
\right).\eeq Next we consider the joint autoregressive
representation of all three processes $X_t$, $Y_t$ and $Z_t$ \be
\label{xtriple} X_t=\sum^{\infty}_{j=1} a_{4j}
X_{t-j}+\sum^{\infty}_{j=1} b_{4j} Y_{t-j}+\sum^{\infty}_{j=1}
c_{4j} Z_{t-j} +\epsilon_{4t}, \ee \be \label{ytriple}
Y_t=\sum^{\infty}_{j=1} d_{4j} X_{t-j}+\sum^{\infty}_{j=1} e_{4j}
Y_{t-j}+\sum^{\infty}_{j=1} g_{4j} Z_{t-j}+\eta_{4t},\ee \be
\label{ztriple} Z_t=\sum^{\infty}_{j=1} u_{4j}
X_{t-j}+\sum^{\infty}_{j=1} v_{4j} Y_{t-j}+\sum^{\infty}_{j=1}
w_{4j} Z_{t-j}+\gamma_{4t},\ee where the covariance matrix of the
noise terms is
\beq \nonumber \mathbf{\Sigma}_4 = \left(%
\begin{array}{ccc}
  \Sigma_{xx} \ \ &   \Sigma_{xy} \ \ &  \Sigma_{xz} \\
    \Sigma_{yx} \ \ &   \Sigma_{yy} \ \ &   \Sigma_{yz} \\
    \Sigma_{zx} \ \ &   \Sigma_{zy} \ \ &   \Sigma_{zz} \\
\end{array}%
\right).  \eeq From these two sets of equations we define the
Granger causality from $Y_t$ to $X_t$ conditional on
$Z_t$\index{conditional Granger causality} to be \be
\label{conditionaltimedomain} F_{Y\rightarrow
X|Z}=\ln\frac{\Sigma_{3}}{\Sigma_{xx}}. \ee The intuitive meaning of
this definition is quite clear. When the causal influence from $Y_t$
to $X_t$ is entirely mediated by $Z_t$ (Fig. 1(a)), $\{b_{4j}\}$ is
uniformly zero, and $\Sigma_{xx}=\Sigma_3$. Thus, we have
$F_{Y\rightarrow X|Z}=0$, meaning that no further improvement in the
prediction of $X_t$ can be expected by including past measurements
of $Y_t$. On the other hand, when there is still a direct component
from $Y_t$ to $X_t$ (Fig. 1(b)), the inclusion of past measurements
of $Y_t$ in addition to that of $X_t$ and $Z_t$ results in better
predictions of $X_t$, leading to $\Sigma_{xx}<\Sigma_3$, and
$F_{Y\rightarrow X|Z}>0$.

\subsection{Frequency Domain Formulation}

To derive the spectral decomposition of the time domain conditional
Granger causality we carry out a normalization procedure like that
for the bivariate case. For Eqs. (\ref{xcdouble}) and
(\ref{zcdouble}) the normalized equations are
\beq \label{xzdouble} \left(%
\begin{array}{cc}
  D_{11}(L) \ \ & D_{12}(L) \\
  D_{21}(L) \ \ & D_{22}(L) \\
\end{array}%
\right)\left(%
\begin{array}{c}
  x_t \\
  z_t \\
\end{array}%
\right) =\left(%
\begin{array}{c}
  x_t^* \\
  z_t^* \\
\end{array}%
\right), \eeq where $D_{11}(0)=1, D_{22}(0)=1, D_{12}(0)=0$,
$\mathrm{cov}(x^*_t,z^*_t)=0$, and $D_{21}(0)$ is generally not
zero. We note that $\rm{var}(x_t^*)=\Sigma_3$ and this becomes
useful in what follows.

For Eqs. (\ref{xtriple}), (\ref{ytriple}) and (\ref{ztriple}) the
normalization process involves left-multiplying both sides by the
matrix \beq \nonumber \mathbf{P}= \mathbf{P}_2 \cdot
\mathbf{P}_1\eeq where
\beq \nonumber \mathbf{P}_1=\left(%
\begin{array}{ccc}
   1 \ \ &   0 \ \ &   0 \\
  - \Sigma_{yx} \Sigma_{xx}^{-1} \ \ &  1 \ \ \ &  0 \\
  - \Sigma_{zx} \Sigma_{xx}^{-1} \ \ & 0 \ \ \ &  1 \\
\end{array}%
\right),\eeq and
\beq \nonumber \mathbf{P}_2=\left(%
\begin{array}{ccc}
  1 \ \ & 0 \ \ & 0 \\
  0 \ \ & 1 \ \ & 0 \\
  0 \ \ & -(\Sigma_{zy}-\Sigma_{zx}\Sigma_{xx}^{-1}\Sigma_{xy})(\Sigma_{yy}-\Sigma_{yx}\Sigma_{xx}^{-1}\Sigma_{xy})^{-1} \ \ &  1\\
\end{array}%
\right).\eeq We denote the normalized equations as
\beq \label{xyz} \left(%
\begin{array}{ccc}
 B_{11}(L) \ \ & B_{12}(L) \ \ & B_{13}(L) \\
  B_{21}(L) \ \ & B_{22}(L) \ \ & B_{23}(L) \\
  B_{31}(L) \ \ & B_{32}(L) \ \ & B_{33}(L) \\
\end{array}%
\right)\left(%
\begin{array}{c}
  x_t \\
  y_t \\
  z_t \\
\end{array}%
\right) =\left(%
\begin{array}{c}
  \epsilon_{xt} \\
  \epsilon_{yt} \\
  \epsilon_{zt} \\
\end{array}%
\right), \eeq where the noise terms are independent, and their
respective variances are $\hat{\Sigma}_{xx}$, $\hat{\Sigma}_{yy}$
and $\hat{\Sigma}_{zz}$.

To proceed further we need the following important relation
\cite{geweke2} \be F_{Y\rightarrow X|Z}=F_{YZ^*\rightarrow X^*}
\ee and its frequency domain counterpart:\be \label{relation}
\textsl{f}_{Y\rightarrow X|Z}(\omega)=\textsl{f}_{YZ^*\rightarrow
X^*}(\omega). \ee To obtain $\textsl{f}_{YZ^*\rightarrow
X^*}(\omega)$, we need to decompose the spectrum of $X^*$. The
Fourier transform of Eqs. (\ref{xzdouble}) and (\ref{xyz}) gives:
\beq \label{xzf} \left(%
\begin{array}{c}
  X(\omega) \\
  Z(\omega) \\
\end{array}%
\right) =\left(%
\begin{array}{cc}
  G_{xx}(\omega) \ \ & G_{xz}(\omega) \\
  G_{zx}(\omega) \ \ & G_{zz}(\omega) \\
\end{array}%
\right)\left(%
\begin{array}{c}
  X^*(\omega) \\
  Z^*(\omega) \\
\end{array}%
\right), \eeq and
\beq \label{xyzf} \left(%
\begin{array}{c}
  X(\omega) \\
  Y(\omega) \\
  Z(\omega) \\
\end{array}%
\right) =\left(%
\begin{array}{ccc}
  H_{xx}(\omega) \ \ & H_{xy}(\omega) \ \ & H_{xz}(\omega) \\
  H_{yx}(\omega) \ \ & H_{yy}(\omega) \ \ & H_{yz}(\omega) \\
  H_{zx}(\omega) \ \ & H_{zy}(\omega) \ \ & H_{zz}(\omega) \\
\end{array}%
\right)\left(%
\begin{array}{c}
  E_x(\omega) \\
  E_y(\omega) \\
  E_z(\omega) \\
\end{array}%
\right). \eeq Assuming that $X(\omega)$ and $Z(\omega)$ from Eq.
(\ref{xzf}) can be equated with that from Eq. (\ref{xyzf}), we
combine Eq. (\ref{xzf}) and Eq. (\ref{xyzf}) to yield,
\beq \label{fdc} \left(%
\begin{array}{c}
  X^*(\omega) \\
  Y(\omega) \\
  Z^*(\omega) \\
\end{array}%
\right) &=& \left(%
\begin{array}{ccc}
  G_{xx}(\omega) \ \ & 0 \ \ & G_{xz}(\omega) \\
  0 \ \ & 1 \ \ & 0 \\
  G_{zx}(\omega) \ \ & 0 \ \ & G_{zz}(\omega) \\
\end{array}%
\right)^{-1}\left(%
\begin{array}{ccc}
  H_{xx}(\omega) \ \ & H_{xy}(\omega) \ \ & H_{xz}(\omega) \\
  H_{yx}(\omega) \ \ & H_{yy}(\omega) \ \ & H_{yz}(\omega) \\
  H_{zx}(\omega) \ \ & H_{zy}(\omega) \ \ & H_{zz}(\omega) \\
\end{array}%
\right)\left(%
\begin{array}{c}
  E_x(\omega) \\
  E_y(\omega) \\
  E_z(\omega) \\
\end{array}%
\right)\nonumber \\
 &=& \left(%
\begin{array}{ccc}
  Q_{xx}(\omega) \ \ & Q_{xy}(\omega) \ \ & Q_{xz}(\omega) \\
  Q_{yx}(\omega) \ \ & Q_{yy}(\omega) \ \ & Q_{yz}(\omega) \\
  Q_{zx}(\omega) \ \ & Q_{zy}(\omega) \ \ & Q_{zz}(\omega) \\
\end{array}%
\right)\left(%
\begin{array}{c}
  E_x(\omega) \\
  E_y(\omega) \\
  E_z(\omega) \\
\end{array}%
\right), \eeq where
$\mbf{Q}(\omega)=\mbf{G}^{-1}(\omega)\mbf{H}(\omega)$. After
suitable ensemble averaging, the spectral matrix can be obtained
from which the power spectrum of $X^*$ is found to be \be
S_{x^*x^*}(\omega)=Q_{xx}(\omega)\hat{\Sigma}_{xx}
Q_{xx}^*(\omega)+Q_{xy}(\omega)\hat{\Sigma}_{yy}
Q_{xy}^*(\omega)+Q_{xz}(\omega)\hat{\Sigma}_{zz}
Q_{xz}^*(\omega).\ee The first term can be thought of as the
intrinsic power and the remaining two terms as the combined causal
influences from $Y$ and $Z^*$. This interpretation leads immediately
to the definition \be \textsl{f}_{YZ^*\rightarrow X^*}(\omega)=\ln
\frac{\left|S_{x^*x^*}(\omega)\right|}{\left|Q_{xx}(\omega)\hat{\Sigma}_{xx}Q_{xx}^*(\omega)\right|}.\ee
We note that $S_{x^*x^*}(\omega)$ is actually the variance of
$\epsilon_{3t}$ as pointed out earlier. On the basis of the relation
in Eq. (\ref{relation}), the final expression for Granger causality
from $Y_t$ to $X_t$ conditional on $Z_t$\index{conditional Granger
causality} is \be \label{cm} \textsl{f}_{Y\rightarrow
X|Z}(\omega)=\ln
\frac{\Sigma_3}{\left|Q_{xx}(\omega)\hat{\Sigma}_{xx}Q_{xx}^*(\omega)\right|}.\ee
It can be shown that $f_{Y\rightarrow X|Z}(\omega)$ relates to the
time domain measure $F_{Y\rightarrow X|Z}$ via \beq F_{Y\rightarrow
X|Z}=\frac{1}{2\pi}\int_{-\pi}^{\pi} f_{Y\rightarrow X|Z}(\omega)d
\omega,\nonumber \eeq under general conditions.

The above derivation is made possible by the key assumption that
$X(\omega)$ and $Z(\omega)$ in Eq. (\ref{xzf}) and in Eq.
(\ref{xyzf}) are identical. This certainly holds true on purely
theoretical grounds, and it may very well be true for simple
mathematical systems. For actual physical data, however, this
condition may be very hard to satisfy due to practical estimation
errors. In a recent paper we developed a partition matrix
technique to overcome this problem \cite{cond}. The subsequent
calculations of conditional Granger causality are based on this
partition matrix procedure.

\section{Estimation of Autoregressive Models}

The preceding theoretical development assumes that the time series
can be well represented by autoregressive processes. Such
theoretical autoregressive processes have infinite model orders.
Here we discuss how to estimate autoregressive models from
empirical time series data, with emphasis on the incorporation of
multiple time series segments into the estimation procedure
\cite{ding}. This consideration is motivated by the goal of
applying autoregressive modeling in neuroscience. It is typical in
behavioral and cognitive neuroscience experiments for the same
event to be repeated on many successive trials. Under appropriate
conditions, time series data recorded from these repeated trials
may be viewed as realizations of a common underlying stochastic
process.

Let ${\bf X}_t=[X_{1t},X_{2t},\cdots,X_{pt}]^T$ be a $p$ dimensional
random process. Here $T$ denotes matrix transposition. In
multivariate neural data, $p$ represents the total number of
recording channels. Assume that the process ${\bf X}_t$ is
stationary and can be described by the following $m$th order
autoregressive equation\index{autoregressive model}
\begin{equation}
{\bf X}_t+{\bf A}(1){\bf X}_{t-1}+\cdots+{\bf A}(m){\bf
X}_{t-m}={\bf E}_t, \label{eq:mvar}
\end{equation}
where ${\bf A}(i)$ are $p \times p$ coefficient matrices and ${\bf
E}_t=[E_{1t},E_{2t},\cdots,E_{pt}]^T$ is a zero mean uncorrelated
noise vector with covariance matrix ${\bf \Sigma}$.

To estimate ${\bf A}(i)$ and ${\bf \Sigma}$, we multiply
Eq.~(\ref{eq:mvar}) from the right by ${\bf X}^T_{t-k}$, where
$k=1,2,\cdots,m$. Taking expectations, we obtain the Yule-Walker
equations
\begin{equation}
{\bf R}(-k)+{\bf A}(1){\bf R}(-k+1)+\cdots+{\bf A}(m){\bf
R}(-k+m)={\bf 0}, \label{eq:yule-walker}
\end{equation}
where ${\bf R}(n)=<{\bf X}_t{\bf X}^T_{t+n}>$ is ${\bf X}_t$'s
covariance matrix of lag $n$. In deriving these equations, we have
used the fact that $<{\bf E}_t{\bf X}^T_{t-k}>={\bf 0}$ as a result
of ${\bf E}_t$ being an uncorrelated process.

For a single realization of the ${\bf X}$ process, $\{{\bf
x}_i\}_{i=1}^N$, we compute the covariance matrix in
Eq.~(\ref{eq:yule-walker}) according to
\begin{equation}
\tilde{\bf R}(n)=\frac{1}{N-n}\sum_{i=1}^{N-n}{\bf x}_i{\bf
x}^T_{i+n}. \label{eq:corr}
\end{equation}
If multiple realizations of the same process are available, then
we compute the above quantity for each realization, and average
across all the realizations to obtain the final estimate of the
covariance matrix. (Note that for a single short trial of data one
uses the divisor $N$ for evaluating covariance to reduce
inconsistency. Due to the availability of multiple trials in
neural applications, we have used the divisor $(N-n)$ in the above
definition Eq.~({\ref{eq:corr}) to achieve an unbiased estimate.)
It is quite clear that, for a single realization, if $N$ is small,
one will not get good estimates of ${\bf R}(n)$ and hence will not
be able to obtain a good model. This problem can be overcome if a
large number of realizations of the same process is available. In
this case the length of each realization can be as short as the
model order $m$ plus 1.

Equations (\ref{eq:mvar}) contain a total of $mp^2$ unknown model
coefficients. In (\ref{eq:yule-walker}) there is exactly the same
number of simultaneous linear equations. One can simply solve
these equations to obtain the model coefficients. An alternative
approach is to use the Levinson, Wiggins, Robinson (LWR)
algorithm, which is a more robust solution procedure based on the
ideas of maximum entropy. This algorithm was implemented in the
analysis of neural data described below. The noise covariance
matrix ${\bf \Sigma}$ may be obtained as part of the LWR
algorithm. Otherwise one may obtain ${\bf \Sigma}$ through
\begin{equation}
{\bf \Sigma}={\bf R}(0)+\sum_{i=1}^m {\bf A}(i){\bf R}(i).
\label{eq:sigma}
\end{equation}
Here we note that ${\bf R}^T(k)={\bf R}(-k)$.

The above estimation procedure can be carried out for any model
order $m$. The correct $m$ is usually determined by minimizing the
Akaike Information Criterion (AIC)\index{Akaike Information
Criterion (AIC)} defined as
\begin{equation}
{\mbox{AIC}}(m)=2\log[\mbox{det}({\bf
\Sigma})]+\frac{2p^2m}{N_{\mbox {total}}} \label{eq:aic}
\end{equation}
where $N_{\mbox{total}}$ is the total number of data points from all
the trials. Plotted as a function of $m$ the proper model order
correspond to the minimum of this function. It is often the case
that for neurobiological data $N_{\mbox{total}}$ is very large.
Consequently, for a reasonable range of $m$, the AIC function does
not achieve a minimum. An alternative criterion is the Bayesian
Information Criterion (BIC)\index{Bayesian Information Criterion
(BIC)}, which is defined as
\begin{equation}
{\mbox{BIC}}(m)=2\log[\mbox{det}({\bf \Sigma})]+\frac{2p^2m\log
N_{\mbox{total}}}{N_{\mbox {total}}} \label{eq:bic}
\end{equation}
This criterion can compensate for the large number of data points
and may perform better in neural applications. A final step,
necessary for determining whether the autoregressive time series
model is suited for a given data set, is to check whether the
residual noise is white. Here the residual noise is obtained by
computing the difference between the model's predicted values and
the actually measured values.

Once an autoregressive model is adequately estimated, it becomes
the basis for both time domain and spectral domain causality
analysis. Specifically, in the spectral domain Eq.~(\ref{eq:mvar})
can be written as
\begin{equation}
{\bf X}(\omega)={\bf H}(\omega){\bf E}(\omega) \label{eq:fourier}
\end{equation}
where
\begin{equation}
{\bf H}(\omega)=(\sum_{j=0}^{m}{\bf A}(j)e^{-i\omega j})^{-1}
\end{equation}
is the transfer function with ${\bf A}(0)$ being the identity
matrix. From Eq.~(\ref{eq:fourier}), after proper ensemble
averaging, we obtain the spectral matrix\index{spectral matrix}
\begin{equation}
{\bf S}(\omega)={\bf H}(\omega){\bf \Sigma}{\bf H}^*(\omega).
\end{equation}
Once we obtain the transfer function, the noise covariance, and
the spectral matrix, we can then carry out causality analysis
according to the procedures outlined in the previous sections.

\section{Numerical Examples}

In this section we consider three examples that illustrate various
aspects of the general approach outlined earlier.

\subsection{Example 1}

Consider the following AR(2) model: \be \label{ar2}
\begin{array}{l}
X_t = 0.9X_{t-1}-0.5X_{t-2}+\epsilon_t \\
Y_t = 0.8Y_{t-1}-0.5Y_{t-2}+0.16X_{t-1}-0.2X_{t-2}+\eta_t\\
\end{array} \ee
where  $\epsilon_t, \eta_t$ are gaussian white noise processes with
zero means and variances $\sigma_1^2=1,\sigma_2^2=0.7$,
respectively. The covariance between $\epsilon_t$ and $\eta_t$ is
0.4. From the construction of the model, we can see that $X_t$ has a
causal influence on $Y_t$ and that there is also instantaneous
causality between $X_t$ and $Y_t$.

We simulated Eq. (\ref{ar2}) to generate a data set of 500
realizations of 100 time points each. Assuming no knowledge of Eq.
(\ref{ar2}) we fitted a MVAR model on the generated data set and
calculated power, coherence and Granger causality spectra. The
result is shown in Figure \ref{figure2}. The interdependence
spectrum is computed according to Eq. (\ref{cohrelation}) and the
total causality is defined as the sum of directional causalities and
the instantaneous causality. The result clearly recovers the pattern
of connectivity in Eq. (\ref{ar2}). It also illustrates that the
interdependence spectrum, as computed according to Eq.
(\ref{cohrelation}), is almost identical to the total causality
spectrum as defined on the right hand side of Eq.
(\ref{freqrelationship}).
\begin{figure}[htb]
\begin{center}
\includegraphics[scale=0.8]{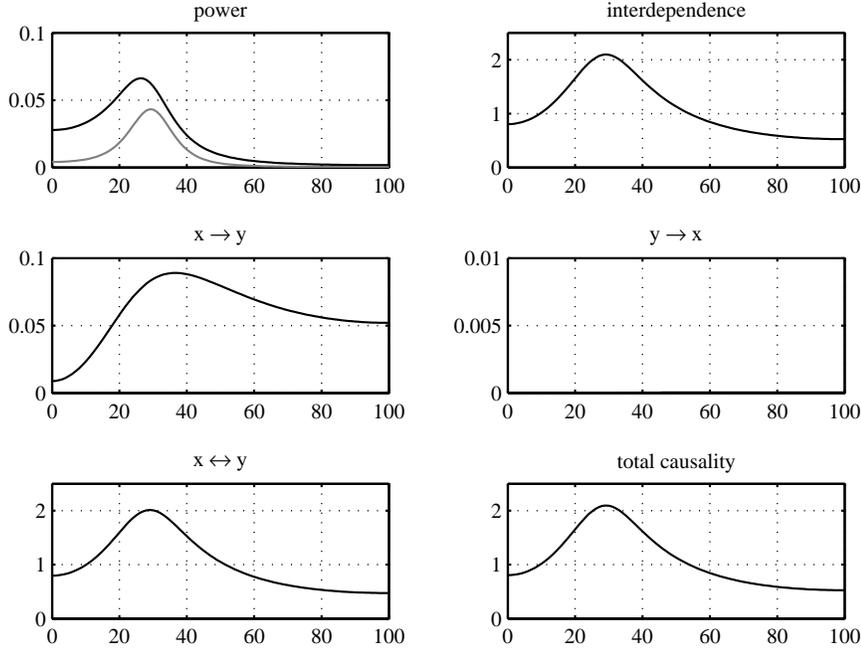}
\end{center}
     \caption{Simulation results for an AR(2) model
consisting of two coupled time series. Power (black for $X$, gray
for $Y$) spectra, interdependence spectrum (related to the coherence
spectrum), and Granger causality spectra are displayed. Note that
the total causality spectrum, representing the sum of directional
causalities and the instantaneous causality, is nearly identical to
the interdependence spectrum.}
     \label{figure2}
\end{figure}

\begin{figure}[htb]
\begin{center}
\includegraphics[scale=0.6]{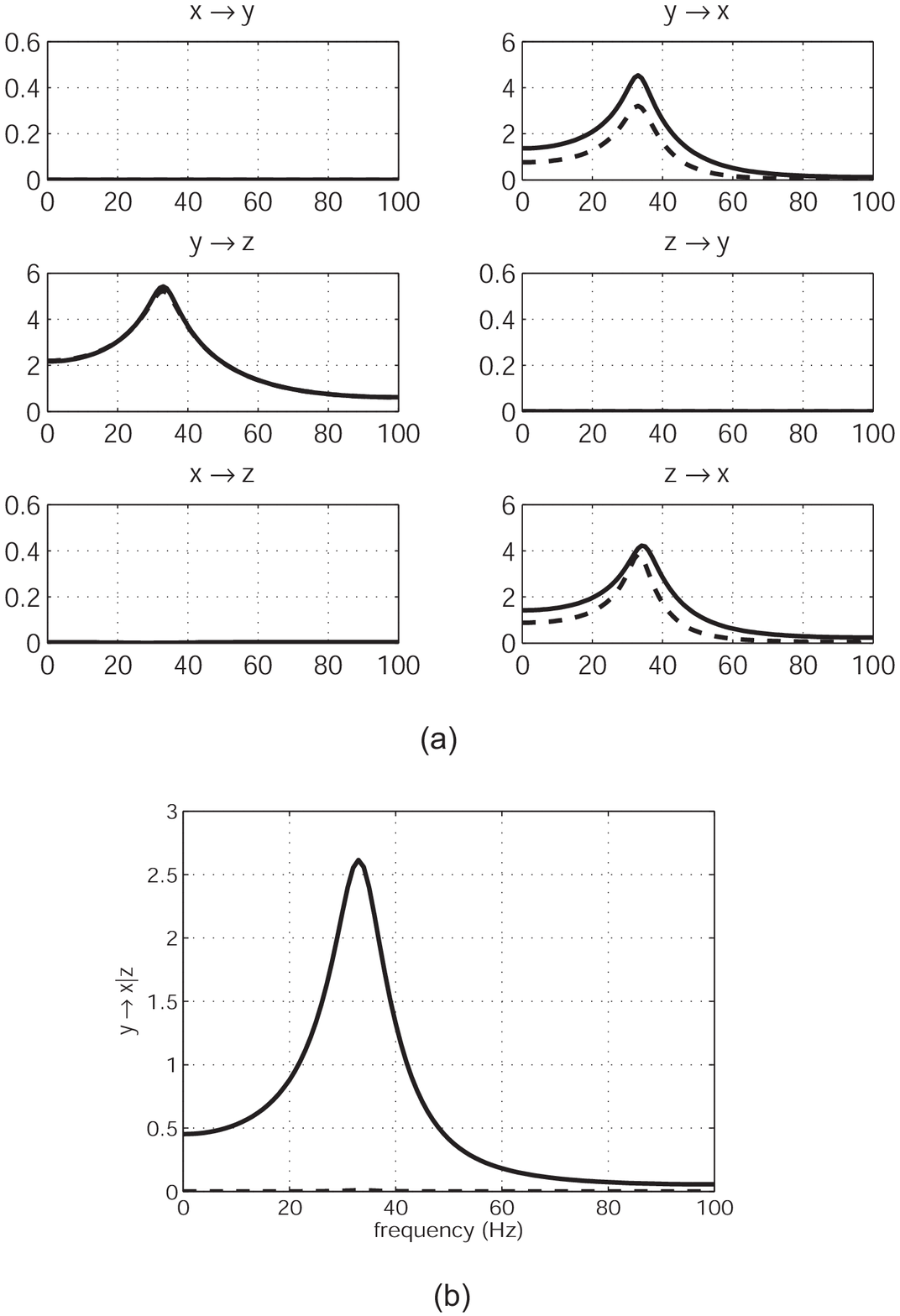}
\end{center}
     \caption{Simulation results for three coupled time
series. Two distinct patterns of connectivity as that illustrated in
Fig. 1 are considered. Results for the case with a direct causal
influence are shown as solid curves and the results for the case
with indirect causal influence are shown as dashed curves. (a)
Pairwise Granger causality analysis gives very similar results for
both cases which indicates that the pairwise analysis cannot
differentiate these two patterns of connectivity. (b) Conditional
causality analysis shows a nonzero spectrum (solid) for the direct
case and almost zero spectrum (dashed) for the indirect case.}
     \label{figure3}
\end{figure}

\subsection{Example 2} Here we consider two models. The first consists of
three time series simulating the case shown in Figure
\ref{figure1}(a), in which the causal influence from $Y_t$ to $X_t$
is indirect and completely mediated by $Z_t$:\be \label{ar3ind}
\begin{array}{l}
X_t = 0.8X_{t-1}-0.5X_{t-2}+0.4Z_{t-1}+\epsilon_t\\
Y_t = 0.9Y_{t-1}-0.8Y_{t-2}+\xi_t \\
Z_t = 0.5Z_{t-1}-0.2Z_{t-2}+0.5Y_{t-1}+\eta_t.\\
\end{array} \ee
The second model creates a situation corresponding to Figure
\ref{figure1}(b), containing both direct and indirect causal
influences from $Y_t$ to $X_t$. This is achieved by using the same
system as in Eq. (\ref{ar3ind}), but with an additional term in the
first equation: \be \label{ar3direct}
\begin{array}{l}
X_t = 0.8X_{t-1}-0.5X_{t-2}+0.4Z_{t-1}+0.2Y_{t-2}+\epsilon_t\\
Y_t = 0.9Y_{t-1}-0.8Y_{t-2}+\xi_t \\
Z_t = 0.5Z_{t-1}-0.2Z_{t-2}+0.5Y_{t-1}+\eta_t.\\
\end{array} \ee
For both models. $\epsilon(t),\xi(t), \eta(t)$ are three independent
gaussian white noise processes with zero means and variances of
$\sigma_1^2=0.3,\sigma_2^2=1,\sigma_3^2=0.2$, respectively.

Each model was simulated to generate a data set of 500 realizations
of 100 time points each. First, pairwise Granger causality analysis
was performed on the simulated data set of each model. The results
are shown in Figure \ref{figure3}(a), with the dashed curves showing
the results for the first model and the solid curves for the second
model. From these plots it is clear that pairwise analysis cannot
differentiate the two coupling schemes. This problem occurs because
the indirect causal influence from $Y_t$ to $X_t$ that depends
completely on $Z_t$ in the first model cannot be clearly
distinguished from the direct influence from $Y_t$ to $X_t$ in the
second model. Next, conditional Granger causality analysis was
performed on both simulated data sets. The Granger causality spectra
from $Y_t$ to $X_t$ conditional on $Z_t$ are shown in Figure
\ref{figure3}(b), with the second model's result shown as the solid
curve and the first model's result as the dashed curve. Clearly, the
causal influence from $Y_t$ to $X_t$ that was prominent in the
pairwise analysis of the first model in Figure \ref{figure3}(a), is
no longer present in Figure \ref{figure3}(b). Thus, by correctly
determining that there is no direct causal influence from $Y_t$ to
$X_t$ in the first model, the conditional Granger causality analysis
provides an unambiguous dissociation of the coupling schemes
represented by the two models.

\subsection{Example 3} We simulated a 5-node oscillatory network structurally
connected with different delays.
\begin{figure}[htb]
\begin{center}
\includegraphics[scale=0.8]{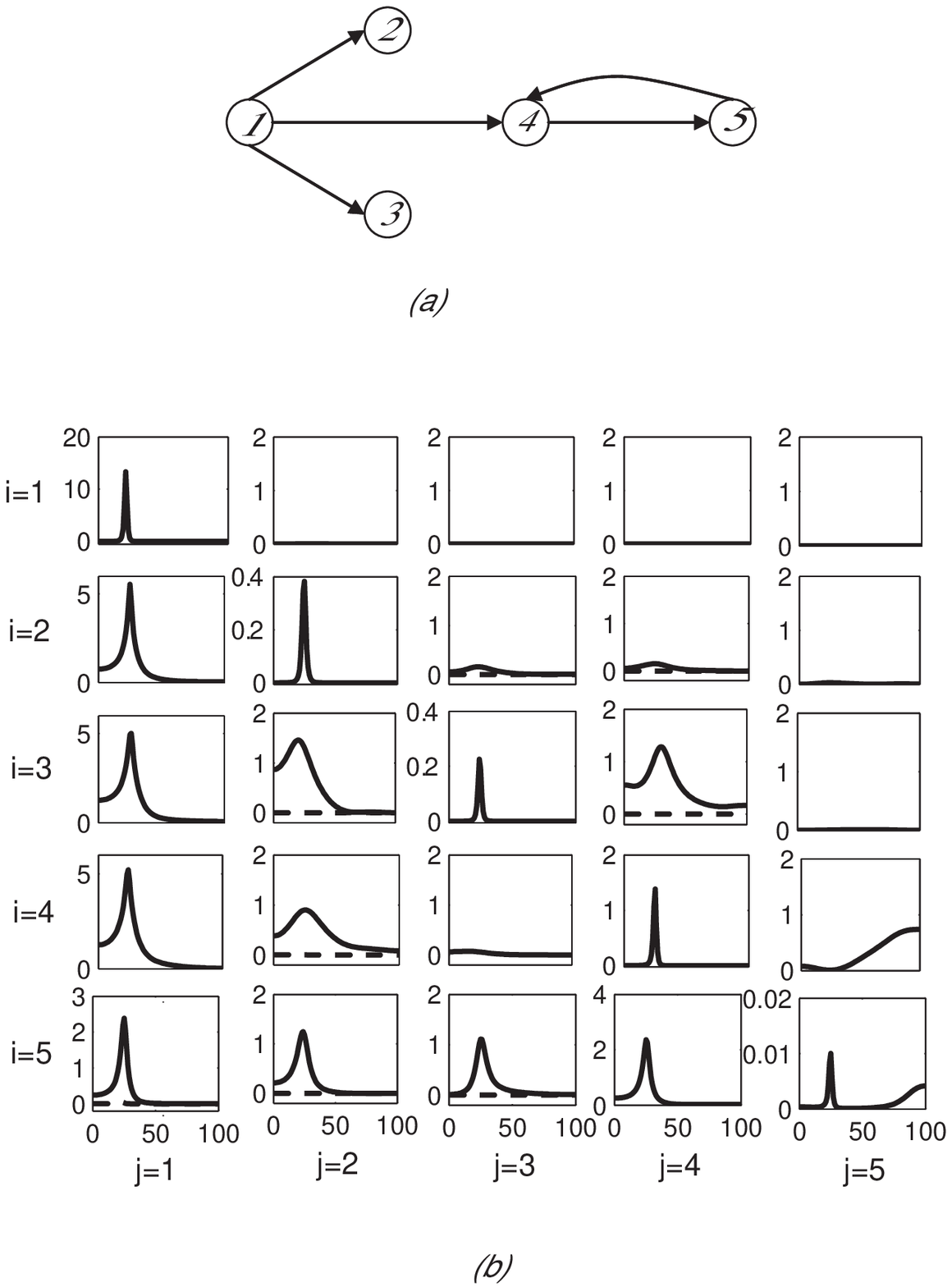}
\end{center}
     \caption{Simulation results for a five-node network structurally
connected with different time delays. (a) Schematic illustration of
the system. (b) Calculated power spectra are shown in the diagonal
panels, results of pairwise (solid) and conditional Granger
causality analysis (dashed) are in off-diagonal panels. Granger
causal influence is from the horizontal index to the vertical index.
Features of Granger causality spectra (both pairwise and
conditional) are consistent with that of power spectra.}
     \label{figure4}
\end{figure}
This example has been analyzed with
partial directed coherence and directed transfer function methods in
\cite{baccala}. The network involves the following multivariate
autoregressive model \be \label{networkmodel}
\begin{array}{l}
X_{1t} = 0.95\sqrt{2}X_{1(t-1)}-0.9025X_{1(t-2)}+\epsilon_{1t}\\
X_{2t} = 0.5X_{1(t-2)}+\epsilon_{2t} \\
X_{3t} = -0.4X_{1(t-3)}+\epsilon_{3t}\\
X_{4t} =
-0.5X_{1(t-2)}+0.25\sqrt{2}X_{4(t-1)}+0.25\sqrt{2}X_{5(t-1)}+\epsilon_{4t}
\\
X_{5t} =
-0.25\sqrt{2}X_{4(t-1)}+0.25\sqrt{2}X_{5(t-1)}+\epsilon_{5t},
\\
\end{array} \ee
where
$\epsilon_{1t},\epsilon_{2t},\epsilon_{3t},\epsilon_{4t},\epsilon_{5t}$
are independent gaussian white noise processes with zero means and
variances of
$\sigma_1^2=0.6,\sigma_2^2=0.5,\sigma_3^2=0.3,\sigma_4^2=0.3,\sigma_5^2=0.6$,
respectively. The structure of the network is illustrated in Figure
\ref{figure4}(a).

We simulated the network model to generate a data set of 500
realizations each with 10 time points. Assuming no knowledge of the
model, we fitted a 5th order MVAR model on the generated data set
and performed power spectra, coherence and Granger causality
analysis on the fitted model. The results of power spectra are given
in the diagonal panels of Figure \ref{figure4}(b). It is clearly
seen that all five oscillators have a spectral peak at around 25Hz
and the fifth has some additional high frequency activity as well.
The results of pairwise Granger causality spectra are shown in the
off-diagonal panels of Figure \ref{figure4}(b) (solid curves).
Compared to the network diagram in Figure \ref{figure4}(a) we can
see that pairwise analysis yields connections that can be the result
of direct causal influences (e.g. $1 \rightarrow 2$), indirect
causal influences (e.g. $1 \rightarrow 5$) and differentially
delayed driving inputs (e.g. $2 \rightarrow 3$). We further
performed a conditional Granger causality analysis in which the
direct causal influence between any two nodes are examined while the
influences from the other three nodes are conditioned out. The
results are shown as dashed curves in Figure \ref{figure4}(b). For
many pairs the dashed curves and solid curves coincide (e.g. $1
\rightarrow 2$), indicating that the underlying causal influence is
direct. For other pairs the dashed curves become zero, indicating
that the causal influences in these pairs are either indirect are
the result of differentially delayed inputs. These results
demonstrate that conditional Granger causality furnishes a more
precise network connectivity diagram that matches the known
structural connectivity. One noteworthy feature about Figure
\ref{figure4}(b) is that the spectral features (e.g. peak frequency)
are consistent across both power and Granger causality spectra. This
is important since it allows us to link local dynamics with that of
the network.

\section{Analysis of a Beta Oscillation Network in Sensorimotor Cortex}

A number of studies have appeared in the neuroscience literature
where the issue of causal effects in neural data is examined
\cite{cond,frei,bern1,bern2,kaminski,baccala,goebel,hesse,andrea}.
Three of these studies \cite{bern1,bern2,andrea} used the measures
presented in this article. Below we review one study published by
our group \cite{cond,andrea}.

Local field potential data were recorded from two macaque monkeys
using transcortical bipolar electrodes at 15 distributed sites in
multiple cortical areas of one hemisphere (right hemisphere in
monkey GE and left hemisphere in monkey LU) while the monkeys
performed a GO/NO-GO visual pattern discrimination task
\cite{bressler}. The prestimulus stage began when the monkey
depressed a hand lever while monitoring a display screen. This was
followed from 0.5 to 1.25 sec later by the appearance of a visual
stimulus (a four-dot pattern) on the screen. The monkey made a GO
response (releasing the lever) or a NO-GO response (maintaining
lever depression) depending on the stimulus category and the
session contingency. The entire trial lasted about 500 ms, during
which the local field potentials were recorded at a sampling rate
of 200 Hz.

Previous studies have shown that synchronized beta-frequency (15-30
Hz) oscillations in the primary motor cortex are involved in
maintaining steady contractions of contralateral arm and hand
muscles. Relatively little is known, however, about the role of
postcentral cortical areas in motor maintenance and their patterns
of interaction with motor cortex. Making use of the simultaneous
recordings from distributed cortical sites we investigated the
interdependency relations of beta-synchronized neuronal assemblies
in pre- and postcentral areas in the prestimulus time period. Using
power and coherence spectral analysis, we first identified a
beta-synchronized large-scale network linking pre- and postcentral
areas. We then used Granger causality spectra to measure directional
influences among recording sites, ascertaining that the dominant
causal influences occurred in the same part of the beta frequency
range as indicated by the power and coherence analysis. The patterns
of significant beta-frequency Granger causality are summarized in
the schematic Granger causality graphs shown in Figure
\ref{figure5}. These patterns reveal that, for both monkeys, strong
Granger causal influences occurred from the primary somatosensory
cortex ($S1$) to both the primary motor cortex ($M1$) and inferior
posterior parietal cortex ($7a$ and $7b$), with the latter areas
also exerting Granger causal influences on the primary motor cortex.
Granger causal influences from the motor cortex to postcentral
areas, however, were not observed\footnote{A more stringent
significance threshold was applied here which resulted in
elimination of several very small causal influences that were
included in the previous report.}.

Our results are the first to demonstrate in awake monkeys that
synchronized beta oscillations not only bind multiple sensorimotor
areas into a large-scale network during motor maintenance behavior,
but also carry Granger causal influences from primary somatosensory
and inferior posterior parietal cortices to motor cortex.
Furthermore, the Granger causality graphs in Figure \ref{figure5}
provide a basis for fruitful speculation about the functional role
of each cortical area in the sensorimotor network. First, steady
pressure maintenance is akin to a closed loop control problem and as
such, sensory feedback is expected to provide critical input needed
for cortical assessment of the current state of the behavior. This
notion is consistent with our observation that primary somatosensory
area ($S1$) serves as the dominant source of causal influences to
other areas in the network. Second, posterior parietal area $7b$ is
known to be involved in nonvisually guided movement. As a
higher-order association area it may maintain representations
relating to the current goals of the motor system. This would imply
that area $7b$ receives sensory updates from area $S1$ and outputs
correctional signals to the motor cortex ($M1$). This
conceptualization is consistent with the causality patterns in
Figure \ref{figure5}. As mentioned earlier, previous work has
identified beta range oscillations in the motor cortex as an
important neural correlate of pressure maintenance behavior. The
main contribution of our work is to demonstrate that the beta
network exists on a much larger scale and that postcentral areas
play a key role in organizing the dynamics of the cortical network.
The latter conclusion is made possible by the directional
information provided by Granger causality analysis.
\begin{figure}[htb]
\begin{center}
\includegraphics[scale=0.8]{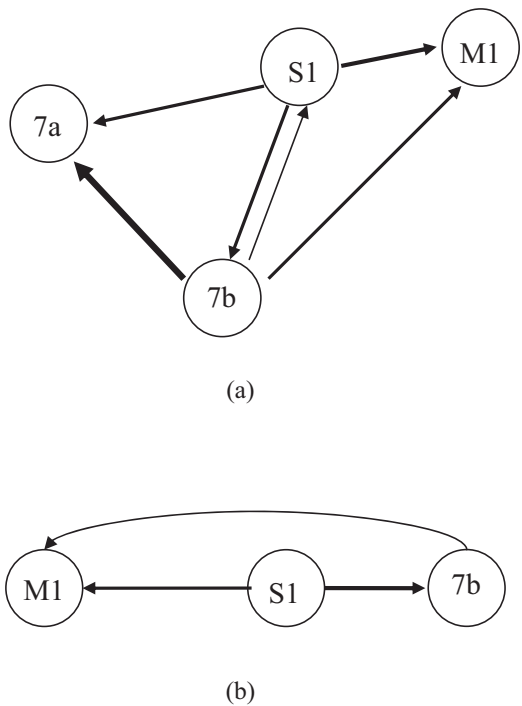}
\end{center}
     \caption{Granger causality graphs for monkey GE (a) and
monkey LU (b).}
     \label{figure5}
\end{figure}

Since the above analysis was pairwise, it had the disadvantage of
not distinguishing between direct and indirect causal influences. In
particular, in monkey GE, the possibility existed that the causal
influence from area $S1$ to inferior posterior parietal area $7a$
was actually mediated by inferior posterior parietal area $7b$
(Figure \ref{figure5}(a)). We used conditional Granger causality to
test the hypothesis that the ${S1 \rightarrow 7a}$ influence was
mediated by area $7b$. In Figure \ref{figure6}(a) is presented the
pairwise Granger causality spectrum from $S1$ to $7a$ (${S1
\rightarrow 7a}$, dark solid curve), showing significant causal
influence in the beta frequency range. Superimposed in Figure
\ref{figure6}(a) is the conditional Granger causality spectrum for
the same pair, but with area $7b$ taken into account (${S1
\rightarrow 7a|7b}$, light solid curve). The corresponding 99\%
significance thresholds are also presented (light and dark dashed
lines coincide). These significance thresholds were determined using
a permutation procedure\index{permutation procedure} that involved
creating 500 permutations of the local field potential data set by
random rearrangement of the trial order independently for each
channel (site). Since the test was performed separately for each
frequency, a correction was necessary for the multiple comparisons
over the whole range of frequencies. The Bonferroni correction could
not be employed because these multiple comparisons were not
independent. An alternative strategy was employed following Blair
and Karniski \cite{blair}. The Granger causality spectrum was
computed for each permutation, and then the maximum causality value
over the frequency range was identified. After 500 permutation
steps, a distribution of maximum causality values was created.
Choosing a p-value at $p=0.01$ for this distribution gave the
thresholds shown in Figure \ref{figure6}(a),(b) and (c) as dashed
lines.

We see from Figure \ref{figure6}(a) that the conditional Granger
causality is greatly reduced in the beta frequency range and no
longer significant, meaning that the causal influence from $S1$ to
$7a$ is most likely an indirect effect mediated by $7b$. This
conclusion is consistent with the known neuroanatomy of the
sensorimotor cortex \cite{fell} in which area $7a$ receives direct
projections from area $7b$ which in turn receives direct projections
from the primary somatosensory cortex. No pathway is known to
project directly from the primary somatosensory cortex to area $7a$.

\begin{figure}[htb]
\begin{center}
\includegraphics[scale=0.6]{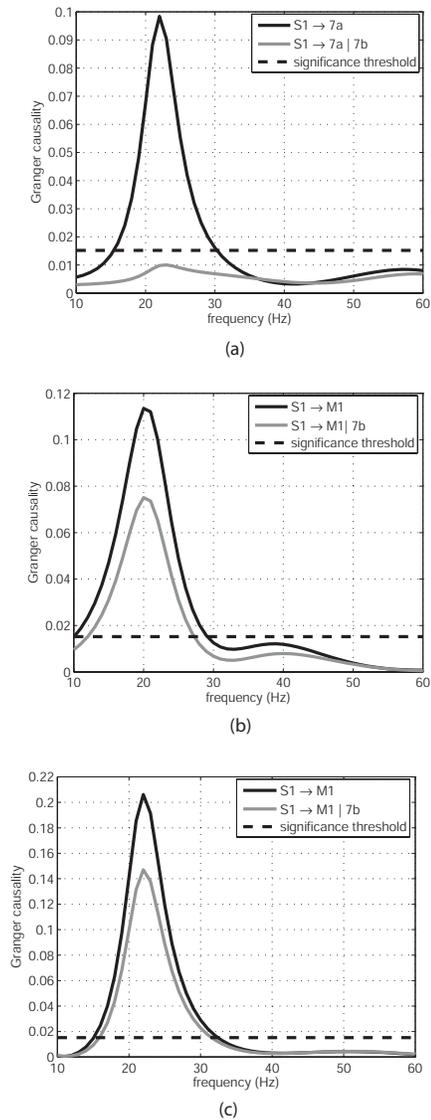}
\end{center}
     \caption{Comparison of pairwise and conditional
Granger causality spectra for monkey GE (a and b), and monkey LU
(c).}
     \label{figure6}
\end{figure}

From Figure \ref{figure5}(a) we see that the possibility also
existed that the causal influence from $S1$ to the primary motor
cortex ($M1$) in monkey GE was mediated by area $7b$. To test this
possibility, the Granger causality spectrum from $S1$ to $M1$ (${S1
\rightarrow M1}$, dark solid curve in Figure \ref{figure6}(b)) was
compared with the conditional Granger causality spectrum with $7b$
taken into account (${S1 \rightarrow M1|7b}$, light solid curve in
Figure \ref{figure6}(b)). In contrast to Figure \ref{figure6}(a), we
see that the beta-frequency conditional Granger causality in Figure
\ref{figure6}(b) is only partially reduced, and remains well above
the 99\% significance level. From Figure \ref{figure4}(b), we see
that the same possibility existed in monkey LU of the $S1$ to $M1$
causal influence being mediated by $7b$. However, just as in Figure
\ref{figure6}(b), we see in Figure \ref{figure6}(c) that the
beta-frequency conditional Granger causality for monkey LU is only
partially reduced, and remains well above the 99\% significance
level.

The results from both monkeys thus indicate that the observed
Granger causal influence from the primary somatosensory cortex to
the primary motor cortex was not simply an indirect effect
mediated by area $7b$. However, we further found that area $7b$
did play a role in mediating the $S1$ to $M1$ causal influence in
both monkeys. This was determined by comparing the means of
bootstrap resampled distributions of the peak beta Granger
causality values from the spectra of ${S1 \rightarrow M1}$ and
${S1 \rightarrow M1|7b}$ by the Student's t-test. The significant
reduction of beta-frequency Granger causality when area $7b$ is
taken into account (t = 17.2 for GE; t = 18.2 for LU, p $<<<$
0.001 for both), indicates that the influence from the primary
somatosensory to primary motor area was partially mediated by area
$7b$. Such an influence is consistent with the known neuroanatomy
\cite{fell} where the primary somatosensory area projects directly
to both the motor cortex and area 7b, and area 7b projects
directly to primary motor cortex.

\section{Summary}

In this article we have introduced the mathematical formalism for
estimating Granger causality in both time and spectral domain from
time series data. Demonstrations of the technique's utilities are
carried out on both simulated data, where the patterns of
interactions are known, and on local field potential recordings from
monkeys performing a cognitive task. For the latter we have stressed
the physiological interpretability of the findings and pointed out
the new insights afforded by these findings. It is our belief that
Granger causality offers a new way of looking at cooperative neural
computation and it enhances our ability to identify key brain
structures underlying the organization of a given brain function.

\end{document}